\documentclass[%
 aip,
 amsmath,amssymb,
 reprint,%
]{revtex4-1}

\usepackage{graphicx}% Include figure files
\usepackage{dcolumn}% Align table columns on decimal point
\usepackage{bm}% bold math
\usepackage{comment}

\usepackage[utf8]{inputenc}
\usepackage[T1]{fontenc}
\usepackage{mathptmx}
\usepackage{etoolbox}
\usepackage[inkscapelatex=false]{svg}

\DeclareUnicodeCharacter{0308}{}

\makeatletter
\def\@email#1#2{%
 \endgroup
 \patchcmd{\titleblock@produce}
  {\frontmatter@RRAPformat}
  {\frontmatter@RRAPformat{\produce@RRAP{*#1\href{mailto:#2}{#2}}}\frontmatter@RRAPformat}
  {}{}
}%
\makeatother
\begin{document}

\preprint{AIP/123-QED}

% inkscape -z -f fig1fab.svg -A fig1fab.pdf
% inkscape fig1fab.svg --export-type=pdf --export-filename=output.pdf

\title[Heterogeneous integration of amorphous silicon carbide on thin film lithium niobate]{Heterogeneous integration of amorphous silicon carbide on thin film lithium niobate}
% Force line breaks with \\
\author{Zizheng Li}
 \email{z.l.li-1@tudelft.nl}
 %Lines break automatically or can be forced with \\
\author{Naresh Sharma}
\author{Bruno Lopez-Rodriguez}
\author{Roald van der Kolk}
\author{Thomas Scholte}
\author{Hugo Voncken}
\author{Jasper van der Boom}
% \author{Silvania F. Pereira}

\affiliation{ 
Department of Imaging Physics (ImPhys), Faculty of Applied Sciences, Delft University of Technology, Delft 2628 CJ, The Netherlands%\\This line break forced with \textbackslash\textbackslash
}%

\author{Simon Gröblacher}

\affiliation{ 
Department of Quantum Nanoscience, Faculty of Applied Sciences, Delft University of Technology, Delft 2628 CJ, The Netherlands%\\This line break forced with \textbackslash\textbackslash
}%

\author{Iman Esmaeil Zadeh}

\affiliation{ 
Department of Imaging Physics (ImPhys), Faculty of Applied Sciences, Delft University of Technology, Delft 2628 CJ, The Netherlands%\\This line break forced with \textbackslash\textbackslash
}%

\date{\today}% It is always \today, today,
             %  but any date may be explicitly specified

\begin{abstract}
In the past decade, lithium niobate (LiNbO$_3$ or LN) photonics, thanks to its heat-free and fast electro-optical modulation, second-order non-linearities and low loss, has been extensively investigated. Despite numerous demonstrations of high-performance LN photonics, processing lithium niobate remains challenging and suffers from incompatibilities with standard complementary metal-oxide semiconductor (CMOS) fabrication lines, limiting its scalability. Silicon carbide (SiC) is an emerging material platform with a high refractive index, a large non-linear Kerr coefficient, and a promising candidate for heterogeneous integration with LN photonics. Current approaches of SiC/LN integration require transfer-bonding techniques, which are time-consuming, expensive, and lack precision in layer thickness. Here we show that amorphous silicon carbide (a-SiC), deposited using inductively coupled plasma enhanced chemical vapor deposition (ICPCVD) at low temperatures (< 165$^\circ$C), can be conveniently integrated with LiNbO$_3$ and processed to form high-performance photonics. Most importantly, the fabrication only involves a standard, silicon-compatible, reactive ion etching step and leaves the LiNbO$_3$ intact, hence its compatibility with standard foundry processes. As a proof-of-principle, we fabricated waveguides and ring resonators on the developed a-SiC/LN platform and achieved intrinsic quality factors higher than 1.06$\times 10^5$ and resonance electro-optic tunability of 3.4 pm/V with 3 mm tuning length. We showcase the possibility of dense integration by fabricating and testing ring resonators with 40$\mu$m radius without a noticeable loss penalty. Our platform offers a CMOS-compatible and scalable approach for implementation of future fast electro-optic modulators and reconfigurable photonic circuits as well as nonlinear processes which can benefit from involving both second and third-order nonlinearities. 
\end{abstract}

\maketitle

% \begin{quotation}
% The ``lead paragraph'' is encapsulated with the \LaTeX\ 
% \verb+quotation+ environment and is formatted as a single paragraph before the first section heading. 
% (The \verb+quotation+ environment reverts to its usual meaning after the first sectioning command.) 
% Note that numbered references are allowed in the lead paragraph.
% %
% The lead paragraph will only be found in an article being prepared for the journal \textit{Chaos}.
% \end{quotation}

% \section{\label{sec:level1}Introduction}

\section{Introduction}

Lithium niobate (LiNbO$_3$ or LN) is a material platform widely utilized in telecommunication as electro-optical modulators, tunable photonic integrated circuits (PICs), and a variety of other applications. Lithium niobate is known for its remarkable properties such as broad transparency window,\cite{zanatta2022optical,han2015optical} low loss,\cite{wu2018lithium,zhu2024twenty,zhuang2023high,desiatov2019ultra} high second order nonlinearity coefficient,\cite{ma2020second,yuan2021strongly,chen2024adapted,wang2019monolithic} strong Pockels effect,\cite{weis1985lithium,thomaschewski2022pockels,boes2023lithium} good physical and chemical stability.\cite{kong2020recent} The platform has been employed in studies in optical communications,\cite{wang2018integrated,li2020lithium,he2019high,boes2023lithium} microwave photonics,\cite{shao2019microwave,boes2023lithium} and quantum computing\cite{maeder2024chip,alibart2016quantum,chapman2023quantum,wang2023quantum,weaver2024integrated}.

To fabricate LiNbO$_3$ PICs, approaches ranging from titanium in-diffusion,\cite{becker1983comparison,chen2022advances} proton exchange,\cite{paz1994characterization,loni1987experimental} direct etching,\cite{zhuang2023high,siew2018ultra,li2023high} dielectric rib-loading circuits,\cite{huang2021high,han2023integrated} wafer bonding\cite{churaev2023heterogeneously,he2019high} and plasmonic waveguides\cite{thomaschewski2020plasmonic} have been successfully demonstrated. The first two methods use bulk LN substrate and locally alter the refractive index in a certain volume to create a small index contrast with respect to the substrate and cladding. These methods can only create weak confinement and hence result in high bending losses and, moreover, they have contamination issues and the fabrication processes are time consuming\cite{ranganath1977suppression}. In contrast, direct etched waveguides and rib-loading circuits based on thin film lithium niobate (TFLN) substrate can overcome these challenges. With the smart-cut technology, a LiNbO$_3$ thin film with a thickness less than 1$\mu$m can be transferred and bonded to an acceptor substrate, which forms the lithium niobate on insulator substrate.\cite{poberaj2009ion,10.1063/5.0037771} Dry etching methods such as reactive ion etching or argon milling etching are commonly chosen to define the PICs on these TFLN films,\cite{10.1063/5.0037771,ulliac2016argon,li2023high} and have already proven high quality and ultralow loss photonic devices.\cite{desiatov2019ultra} However, direct LN etching also has a number of disadvantages: low selectivity between LN and etching masks, non-vertical sidewall angle (typically around 60$^\circ$), re-deposition of etching byproducts such as LiF which is difficult to be removed at low temperature are examples of these issues. In addition, and crucially, etching of LN is not compatible with complementary metal oxide semiconductor (CMOS) fabrication lines, since the etching byproducts and lithium out-diffusion are considered as contamination\cite{chen2022advances,zhu2021integrated}. All of these problems are limiting the possibilities in design and hampering the way towards large scale fabrication and mass production. 

Hybrid integration of rib-loading waveguides with TFLN offers another feasible path to fabricate compact, low-loss, and scalable PICs \cite{han2023integrated}.
%The most significant advantage of this route is that there is no direct LN etching involved, which maintains CMOS compatibility and possibly . 
The integration can be realized using two different schemes, namely, transfer/bond of a pre-fabricated PIC wafer/chip onto a TFLN substrate, or monolithical deposition of a thin film directly on the TFLN substrate followed by fabrication of the rib-loading waveguides on this layer. Regarding the former, various techniques are used to construct the heterogeneous platforms\cite{churaev2023heterogeneously,he2019high,chen2014hybrid,vandekerckhove2023reliable,krishna2024hybrid}. Challenges from transfer bonding process include layer-to-layer misalignment, thermal stress mismatch, and surface non-uniformity and roughness.\cite{mookherjea2023thin,shekhar2024roadmapping,billah2018hybrid,vanackere2023heterogeneous} To circumvent the mentioned disadvantages associated with transfer bonding, delicate and costly procedures such as ion-slicing and multiple rounds of chemical mechanical polishing (CMP) have been utilized, which adds further complexities and limitations.
%Silicon nitride (SiN) rib-loading PIC is a well-studied example of this routine. SiN has a refractive index lower than LN, which results in a large fraction of mode distributed in LN layer and strong interaction between them (...cite SiN/LN nonlinearity also here...). Nevertheless, the ultra-low loss  advantage of SiN can only be leveraged by high temperature processes including low pressure chemical vapor deposition (LPCVD, typically at $800 ^\circ C$) and annealing (typically $1200 ^\circ C$ for nealy 10 hours), implying that.%
In contrast, monolithic deposition and etching scheme is a comparatively more stable and convenient option. Various materials, ranging from amorphous silicon (a-Si) to Ta$_2$O$_5$ and TiO$_2$ have been deposited on TFLN and heterogeneous PIC devices on these platforms have been demonstrated.\cite{cao2014hybrid,chang2016thin,jin2019mid,nakanishi2010high,rabiei2013heterogeneous} Among them, a-Si on TFLN platform offers the merits of small footprint and large thermal-optics tunability but suffers from narrow bandgap and weak mode interaction with LiNbO$_3$, while the others, in addition to similar issues as a-Si, also suffer from incompatibility with CMOS foundries or immature fabrication process.

Amorphous silicon carbide (a-SiC) has recently emerged as a promising option for PICs thanks to its strong optical confinement, wide transparent window, high thermo-optical coefficient, low loss, and high Kerr nonlinear coefficient\cite{lopez2023high,lukin2020integrated,castelletto2022silicon,lukin20204h,xing2020high,xing2019cmos,chang2022demonstration}. Compared to the different polytypes of crystalline silicon carbide used in photonics (3C-SiC, 4H-SiC or 6H-SiC), amorphous silicon carbide material properties can be easily tuned, allows for precise control in layer thickness and has ten times higher Kerr nonlinear coefficient, providing broader possibilities in the field of nonlinear optics. Notably, a-SiC can be deposited at low temperature ( $ \leq 150 ^\circ C $ ) using inductively coupled plasma chemical vapour deposition (ICPCVD) without sacrificing the film quality, which indicates that lift-off of a-SiC films and monolithical integration of devices with different thicknesses and on different platforms are possible.\cite{lopez2023high} Essentially, the a-SiC rib-loading scheme can fulfill all the merits expected for the heterogeneous integration with TFLN, which enhances a multitude of possible applications, including compact and ultra-fast optical modulators, high density and low loss passive PICs,  second order nonlinearity and Kerr nonlinearity, high efficiency on-chip spontaneous down conversion and deterministic integration of single photon sources for quantum photonic applications \cite{zadeh2016deterministic}.
% , monolithic combination of, to name a few. 

% For the first time, the aSiC on TFLN platform is reported, 
In this paper, for the first time a-SiC/TFLN heterogeneous photonic integration is proposed and investigated. Based on an optimized ICPCVD a-SiC on TFLN fabrication processes, we designed, fabricated and characterized on-chip photonic devices and analyzed their performance.  Our results demonstrate the high potentials of the proposed platform in future PIC applications.

\begin{comment}
the optical loss to be 4.5 $\text{dB}/\text{cm}$. The EO tunability of this heterogeneous platform is examined on ground-signal-ground (G-S-G) configured racetrack ring resonators, which gives a resonance tunability of 3.4 pm/V, corresponding to a half-wave voltage-length product result of V$_{\pi}$L = 9.79 $\text{V}\cdot\text{cm}$. Importantly, this method does not incorporate LN etching, wafer bonding, or CMP processes and the fabrication can be done at low temperatures (around $150 ^\circ C$), keeping the CMOS-compatibility and fabrication simplicity
\end{comment}

\section{Methods and results}
\subsection{Fabrication}

\begin{figure*}
    \centering
    \includegraphics[scale=0.17]{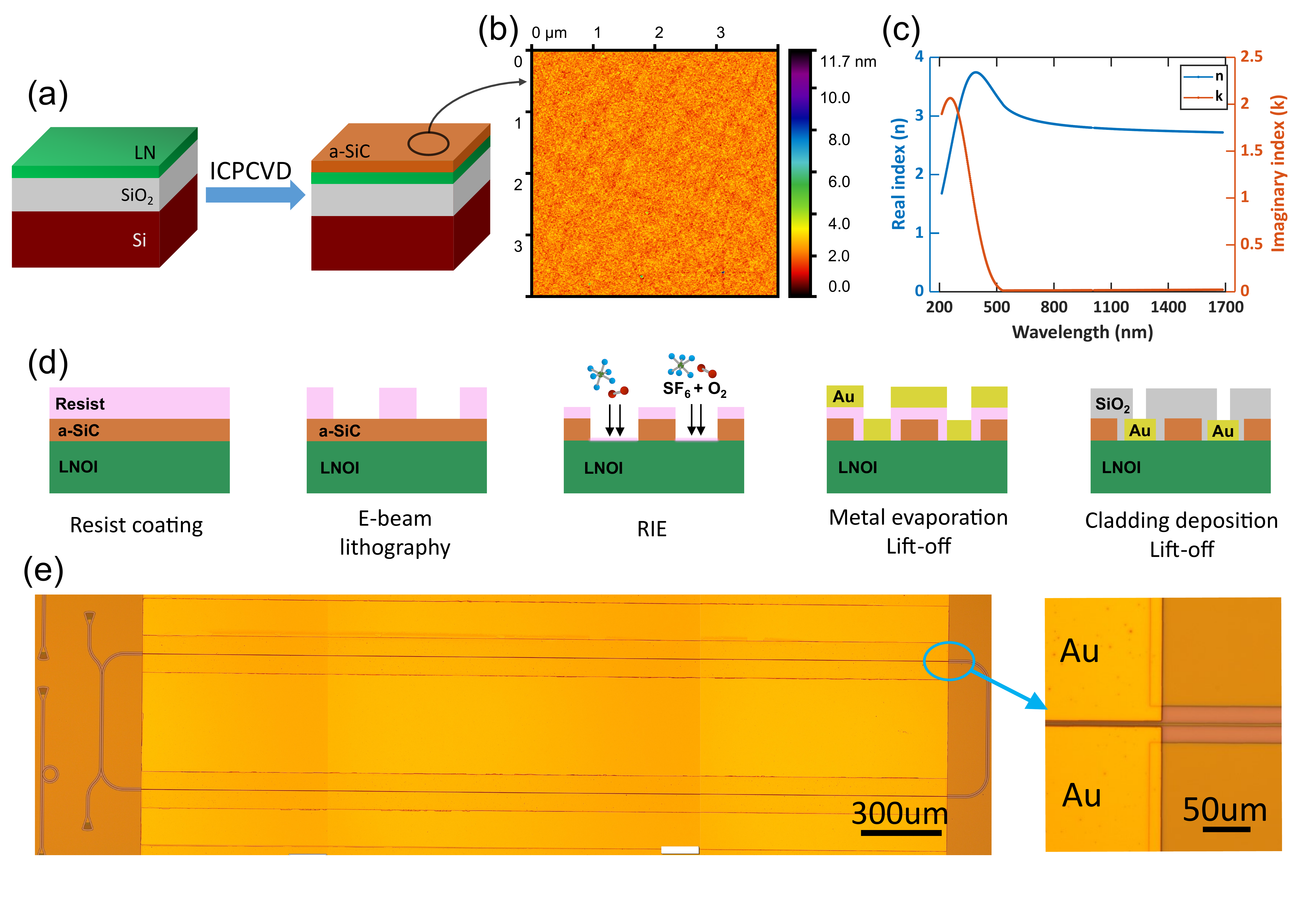}
    \caption{\label{fig:epsart} (a) a-SiC thin film deposition on LNOI substrate. (b) AFM measurement result on a-SiC surface. (c) Real and imaginary refractive index ($n$ and $k$) of the a-SiC film. (d) Fabrication flow of electro-optical tunable a-SiC/TFLN devices. (e) Microscopy image (stiched) of fabricated racetrack ring resonator, with the inset: region of the waveguide between two Au electrodes.}
\end{figure*}

We fabricated the a-SiC rib-loading PICs on the LNOI substrates (NanoLN), which consists of 500 nm TFLN on 2$\mu$m buried SiO$_2$ layer on 300$\mu$m Si substrate. As shown in Fig.1(a), 260 nm a-SiC deposition was done via ICPCVD in Oxford PlasmaPro 100 with mixed precursors of SiH$_4$ and CH$_4$ at a table temperature of 150 $^\circ C$. Argon (Ar) flow was chosen as the deposition environment. The chamber pressure is set to 2 mTorr. It is known that high temperature processes can potentially cause problems, such as unwanted diffusion, non-releasable thermal stress or thermal expansion mismatch, which renders the fabrication process CMOS incompatible. The low-temperature deposited ICPCVD a-SiC film mitigates these challenges, and allows monolithic integration of PICs with different thicknesses on TFLN\cite{lopez2023high}. After a-SiC deposition, ellipsometry measurements were performed (Woollam M-2000 spectroscopic ellipsometer) to find out the refractive index of a-SiC . The real and imaginary index ($n$ and $k$) results are shown in Fig.1(c). Around 1550 nm, wavelength of interest in this work, the complex refractive index is found to be $2.73+\text{i}8.4\times 10^{-5}$ (fitted by B-spline expanded Cauchy model with Cody-Lorentz oscillators\cite{cushman2016introduction}). To determine the surface roughness, which places significant impact on the optical waveguide loss, we conducted an atomic force microscopy (AFM) measurement in a $ 4 \mu m \times 4\mu m$ region on the top surface of a-SiC film (shown in Fig.1(b)). We obtained an root-mean-square (RMS) value of (0.45$\, \pm \,$0.06) nm, indicating good deposition quality and surface flatness. 
%More over, such low surface roughness makes this platform possible to be further heterogeneously integrated with other material platforms by the methods of direct deposition, wafer-bonding, transfer printing, and so on. 

 Fabrication flow for electro-optically tunable devices is shown in Fig.1(d). To fabricate the PIC waveguides, electron beam lithography (EBL) resist (CSAR 62) was spin-coated and baked at 160 $^\circ C$ for 2 minutes. The rib-loading PIC pattern was defined on the resist layer by EBL exposure (Raith 100-kV EBPG-5200), and after development it was transferred to the a-SiC layer by reactive-ion etching (RIE, Sentech Etchlab 200) using a gas mixture of SF$_6$ and O$_2$. The etching time was carefully controlled to prevent over-etching. The electrodes for electro-optics are fabricated by standard lift-off process. 10 nm chromium (Cr) and 450 nm gold (Au) are deposited by electron beam evaporation on the openings, pre-defined by another EBL step, followed by the lift-off process in resist stripper (PRS-3000). Since ICPCVD SiO$_2$ can also be deposited at low temperature (150$^\circ C$), so that compatible to standard lift-off process, the same EBL/lift-off procedure is employed to fabricate the SiO$_2$ cladding. The SiO$_2$ cladding covers all the optical devices but leaving the Au electrodes exposed for the convenience of wire-bonding. Following each EBL exposure, the resist was reflowed at 130 $^\circ C$ for one minute after development, in order to enhance the etching aspect-ratio and reduce the sidewall roughness. It is essential to further clean the resist residuals using oxygen plasma after every chemical lift-off process, to maintain the surface roughness given by ICPCVD deposition and reduce optical scattering brought by remaining particles on the surface. A stitched optical microscopy image of fabricated devices is shown in Fig.1(e), with an inset zooming in on the region of the waveguide sandwiched between the Au electrodes.

\subsection{Design and Simulation}

In the heterogeneous a-SiC/TFLN platform, optical modes are confined and propagated around the rib-loading a-SiC waveguides, in a way that only fundamental TE/TM modes are supported and the mode overlapping fraction with the two materials can be engineered by altering the waveguide width. As illustrated in Fig.2(a), the racetrack ring resonator consists of normal waveguides (width $w_1$ = 800 nm, height $h$ = 260nm), adiabatic tapers and narrow waveguides (width $w_2$ = 400nm, height $h$ = 260 nm). Here we define the mode overlapping fraction in the LN layer as $\Gamma_{\text{LN}}$, denoting the effective percentage of mode intensity distribution and interaction with the material LN. In the standard waveguides, optical modes are relatively strongly confined by the a-SiC waveguides, and the mode overlapping fraction of the fundamental modes are $\Gamma_{\text{LN-TE}} = 33.65\%$ and $\Gamma_{\text{LN-TM}} = 57.04\%$, as shown in Fig.2(b) and (c). Correspondingly, the confinement becomes weaker when the waveguides width is tapered down, which results in $\Gamma_{\text{LN-TE}} = 66.03\%$ and $\Gamma_{\text{LN-TM}} = 68.11\%$ respectively in the narrow waveguides (shown in Fig.2(d) and (e)). In both cases, the waveguides dimensions are designed to ensure that only fundamental modes are supported. The standard waveguides are used for propagation and sharp bendings, utilizing the strong confinement property, while the narrow waveguides can be used for larger interaction between propagation modes and LN layer to enhance EO tuning efficiency. The longer arms of the racetrack ring, marked in Fig.2(a) with $L_2$ = 3000 $\mu$m, are aligned perpendicular to the LN crystal z-axis (marked at the left-bottom corners in Fig.2(b)-(e)) to excite the Pockels effect with the largest EO coefficient $r_{33}$ = 30.9 pm/V \cite{zhu2021integrated}. The mode distributions and overlapping fractions are simulated by finite difference eigenmode solver (MODE, Ansys Lumerical) based on the material properties mentioned in the previous section, taking the refractive index of LN as $n_{\text{e}}$ = 2.13 and $n_{\text{o}}$ = 2.21 around 1550 nm\cite{han2015optical,chen2022advances}. Mesh grid size is chosen close to the lithography resolution and convergence test is performed with perfect matching layer (PML) boundary conditions. Adiabatic tapers with a length of $L_3$ = 100 $\mu$m are adopted to convert the propagation modes between the standard and narrow waveguides. The gap between the bus and ring waveguides is set to 400 nm, and accompanied with the coupling length (shorter arm) $L_1$ = 500 $\mu$m the racetrack ring resonator is brought to critical coupling regime. At the both ends of the bus waveguide, there are two apodized grating couplers enabling efficient in-and-out fiber coupling.

For EO tuning, we use G-S-G electrodes configuration to minimize the tuning length and device footprint, as depicted in Fig.2(a). In order to optimize the tuning efficiency and electrodes parameters, the Charge Transport (CHARGE, Ansys Lumerical) and Finite Element EigenMode (FEEM, Ansys Lumerical) simulations are implemented. Except for the mentioned waveguides geometry, the gap $g$ between the ground and signal is another vital variable that influences the EO tuning performance. There is an inevitable trade-off between higher EO tuning efficiency (smaller $g$) and low propagation loss (larger $g$), which suggests that a balanced point of $g$ needs to be engineered. The propagation loss in the simulations is obtained from the imaginary part of the mode effective index $k$. In the plane where extraordinary LiNbO$_3$ crystal axis is parallel aligned to the applied electric field $E$ (this work focus on this situation mainly), the EO tuning efficiency can be evaluated by the half-wave voltage-length product $V_{\pi} \cdot L$ as: 
$$V_{\pi} \cdot L = \frac{\lambda d}{2n_e^3r_{33}}$$ where d is the distance between the anode and cathode given the electric field $E = V/d$.\cite{zhu2021integrated,churaev2023heterogeneously,krishna2024hybrid} 

Several sets of parameters are compared for both TE and TM modes: $w_1$ = 800 nm, $w_2$ = 400 nm corresponding to the standard waveguide width and narrow waveguide width; signal-ground electrodes gap $g_1$ = 3.8 $\mu m$, $g_2$ = 4.8 $\mu m$, $g_3$ = 4.4 $\mu m$, $g_4$ = 6 $\mu m$. Fig.2(f) shows the effective index change corresponding to the voltage change applied to the signal pad. Clear polarization dependence can be seen, that for fundamental TE mode the effective index change slope is larger than that of the TM mode. This is due to the difference of EO coefficients along the ordinary and extraordinary axis ($r_{33}$ = 30.9 pm/V, $r_{13}$ = 9.6 pm/V). The narrowed waveguide structure has a larger fraction of the optical mode concentrated in the LN layer, resulting in larger index changing slopes. Regarding the propagation losses compared in Fig.2(g), TE polarized modes, for a given gap, have lower losses. At the focused wavelength of 1550\,nm, the absorption from the waveguide materials can be neglected. Consequently, the metal absorption caused by the Au electrodes is the primary contribution to the total propagation loss. From Fig.2(g), it is concluded that for normal waveguides, the metal absorption can be neglected (0.005 dB/cm) when $g\geq$ 4.8 $\mu m$, while for narrow waveguides, $g\geq$ 6 $\mu m$ leads to a relatively larger loss (0.035 dB/cm). Half-wave voltage-length production is simulated and summarized in Fig.2(h). Considering the results in Fig.2(f)-(h), two preferable balanced combinations for TE mode can be found: for normal waveguides $w_1$ = 800 nm, $g_1$ = 4.8 $\mu m$, V$_\pi \cdot L$ = 8.8 V $\cdot$ cm; for narrow waveguides $w_2$ = 400 nm, $g_4$ = 6 $\mu m$, V$_\pi \cdot L$ = 4.7 V $\cdot$ cm, respectively. It allows sharp bends hence dense integration (strong confinement in the normal waveguides) and high electro-optical tunability (weak confinement in the narrow waveguides) on the same chip at the same time. To show the principle while eliminate the potential impact of metal absorption losses, we choose the normal waveguides design to characterize the passive and active response of the devices.

% Here we define the mode overlapping fraction in the LN layer as $\Gamma$, denoting the effective percentage of mode intensity distribution and interaction with the material LN. For the quasi-TE mode (TE polarization fraction $99.1\%$), $\Gamma_{\text{TE}}=33.65\%$; while that of the quasi-TM mode (TM polarization fraction $1.5\%$) is $\Gamma_{\text{TM}}=57.04\%$ in the standard waveguides. 

\begin{figure*}
    \centering
    \includegraphics[scale=0.13]{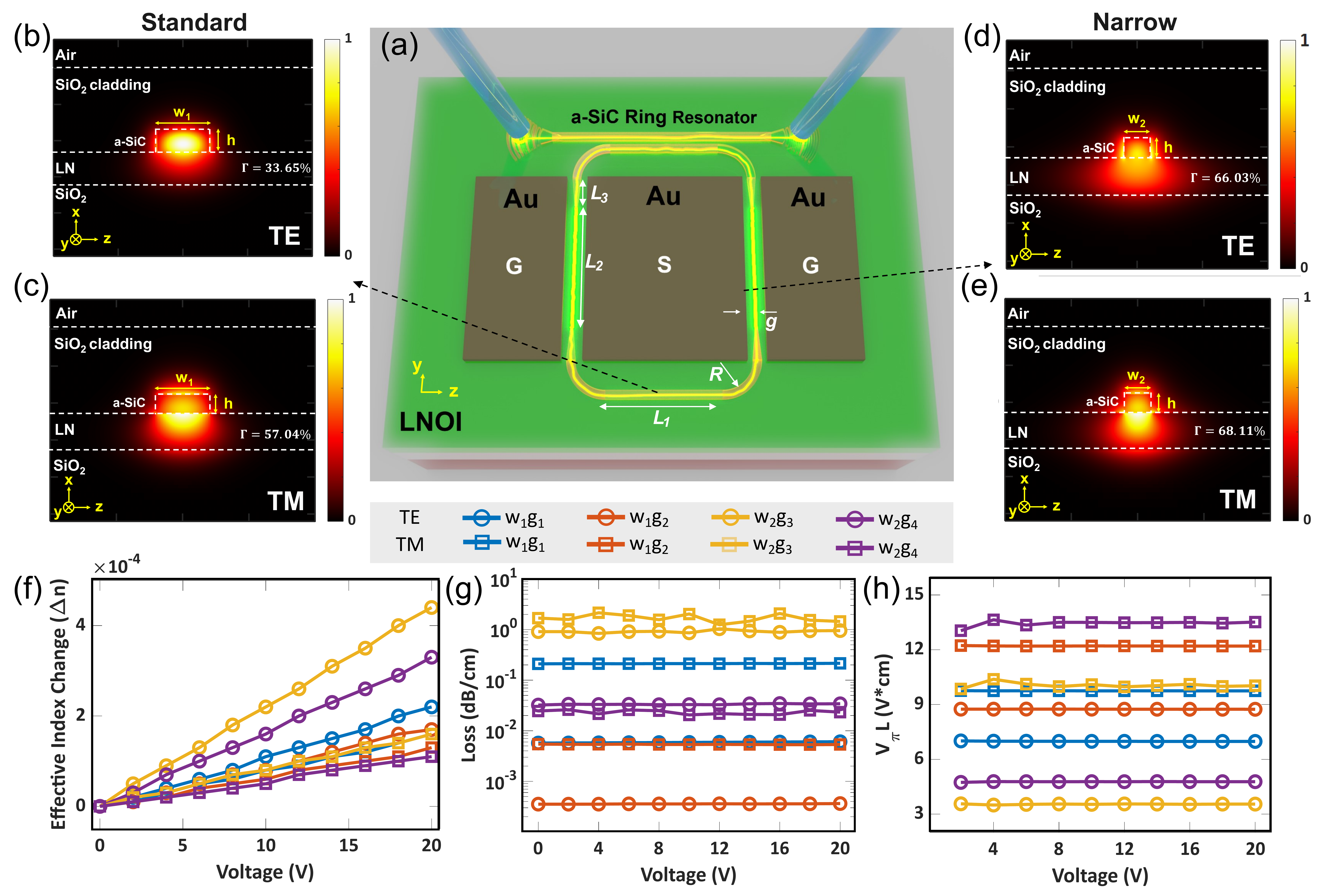}
    \caption{\label{fig:epsart} (a) Schematic of the electro-optical tunable racetrack ring resonator. Mode distribution and mode-LN overlapping fraction of fundamental (b) TE and (c) TM modes in normal waveguides. (d) TE and (e) TM modes distribution with overlapping fraction in tappered down waveguides. (f) Effective index change, (g) propagation loss, and (h) half-wave voltage-length product with respect of applied voltages. (f)-(h) share the same legends, in which w$_1$ = 800 nm, w$_2$ = 400 nm, g$_1$ = 3.8 $\mu$m, g$_2$ = 4.8 $\mu$m, g$_3$ = 4.4 $\mu$m, g$_4$ = 6 $\mu$m, h = 260 nm, and round markers represent TE mode while square markers represent TM mode. }
\end{figure*}

\subsection{Characterization of passive optical devices}

\begin{figure*}
    \centering
    \includegraphics[scale=0.13]{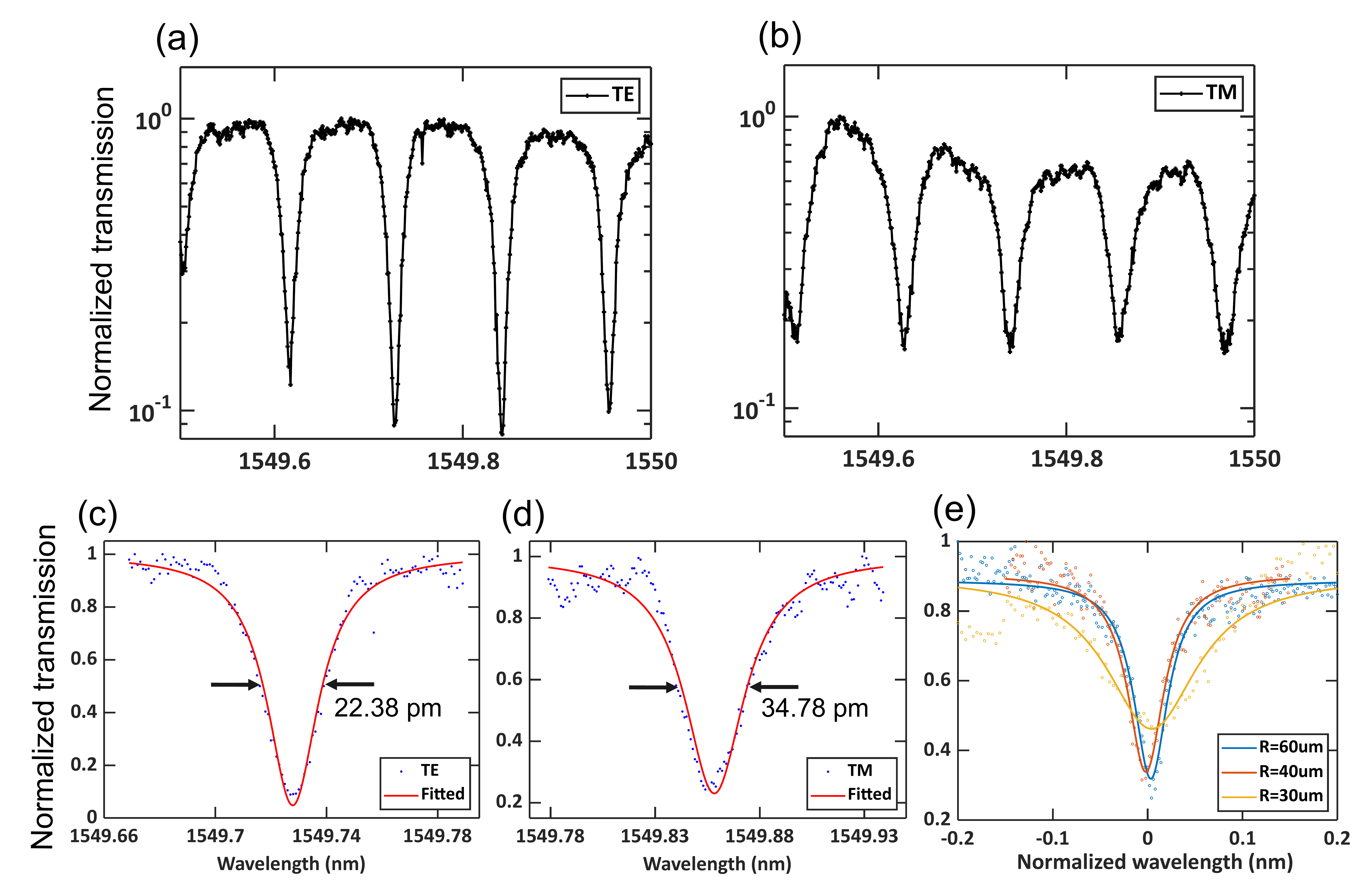}
    \caption{\label{fig:epsart} Characterization of optical properties in the standard waveguide ring resonators: Normalized transmission spectra (dots, log scale) of (a) fundamental TE mode and (b) TM mode. Lorentz fitted resonance dips of (c) TE mode with $FWHM = 22.38$ pm and (d) TM mode with $FWHM = 34.78$ pm. (e) Lorentz fitted resonance dips of ring resonators with different bending radii. }
     % of 60 $\mu$m, 40 $\mu$m, 30 $\mu$m
\end{figure*}

Thin film quality and waveguide propagation loss were characterized using ring resonator with intrinsic quality factor $\text{Q}_{\text{i}}$,\cite{lopez2023high} which is defined as: 
$$ Q_{\text{i}} = \frac{2 Q_{\text{L}}}  {1 + \sqrt{T}} = \frac{2 \lambda_0}  { FWHM (1 + \sqrt{T}) } $$ where $Q_L$ is the loaded quality factor, $\lambda_0$ is the resonance wavelength, $FWHM$ represents the full width half maximum of the resonance dip, and $T$ is the on-resonance transmission. To extract the propagation loss from the intrinsic quality factor, we calculate loss $\alpha$ as: 
$$\alpha \, \text{[dB/cm]} = 10lg_e(\frac{2\pi n_g}{Q_i\lambda_0})$$ where the group index $n_g$ is derived from free spectral range (FSR) of the optical cavity:
$$n_g = \frac{\lambda_0^2}{L_{total} \cdot FSR} = \frac{\lambda_0^2}{2(\pi R + L_2 + L_1) \cdot FSR}$$ in which $L_{Total}$ represents the total length of the ring.
The $FSR$ can be obtained from the resonance spectra shown in Fig.3(a) and (b). In comparison to the simulation results mentioned in the previous section, for a ring consisting of normal waveguides (260 nm $\times$ 800nm), the simulated group indices for TE and TM modes are $n_{g\text{TE}}$ = 2.738 and $n_{g\text{TM}}$ = 2.614 at 1550 nm, corresponding to $FSR$ of 113 pm and 118 pm respectively, which are in good agreement with the experimental results. 
%For TE&TM ng should be different, then FSR should also be different. In simulation, ngTE=2.73849+1.4e-07i; ngTM=2.61396+5.8e-07i. So FSR_TE=lambda^2/L/ngTE=113pm, FSR_TM=118pm. Correct! ...
Single resonance peaks acquired from the TE and TM spectra are fitted by the Lorentzian function (Fig.3(c) and (d)), showing $FWHM$ of 22.38 pm for TE mode and 34.78 pm for TM mode, respectively. The calculated intrinsic quality factor for TE mode is 106,673, denoting a propagation loss of 4.48 dB/cm. 

Ring resonators with different bending radii are fabricated and measured to analyze the influence of bending loss. Here, the gap between ring and bus waveguide is not critical, as the only focus is on the change of $FWHM$ representing the variation in loss, regardless it is in the under-coupling or over-coupling regime. The Lorentzian fitting of the resonance dips from different rings are brought together in Fig.2(f) for comparison, in which one can see when decreasing the bending radius the $FWHM$ does not change until $R$=40$\mu$m, while the $FWHM$ becomes much wider from $R$=40$\mu$m to $R$=30$\mu$m. It is inferred that the bending loss can be neglected for $R\geq$40$\mu$m, which enables denser integration compared with SiN rib-loading waveguides. Nevertheless, in the following context that we discuss electro-optics tuning, the bending radius of racetrack ring resonators is kept to 120 $\mu$m, completely ruling out any bending losses.

\subsection{Characterization of EO tunability}

\begin{figure*}
    \centering
    \includegraphics[scale=0.17]{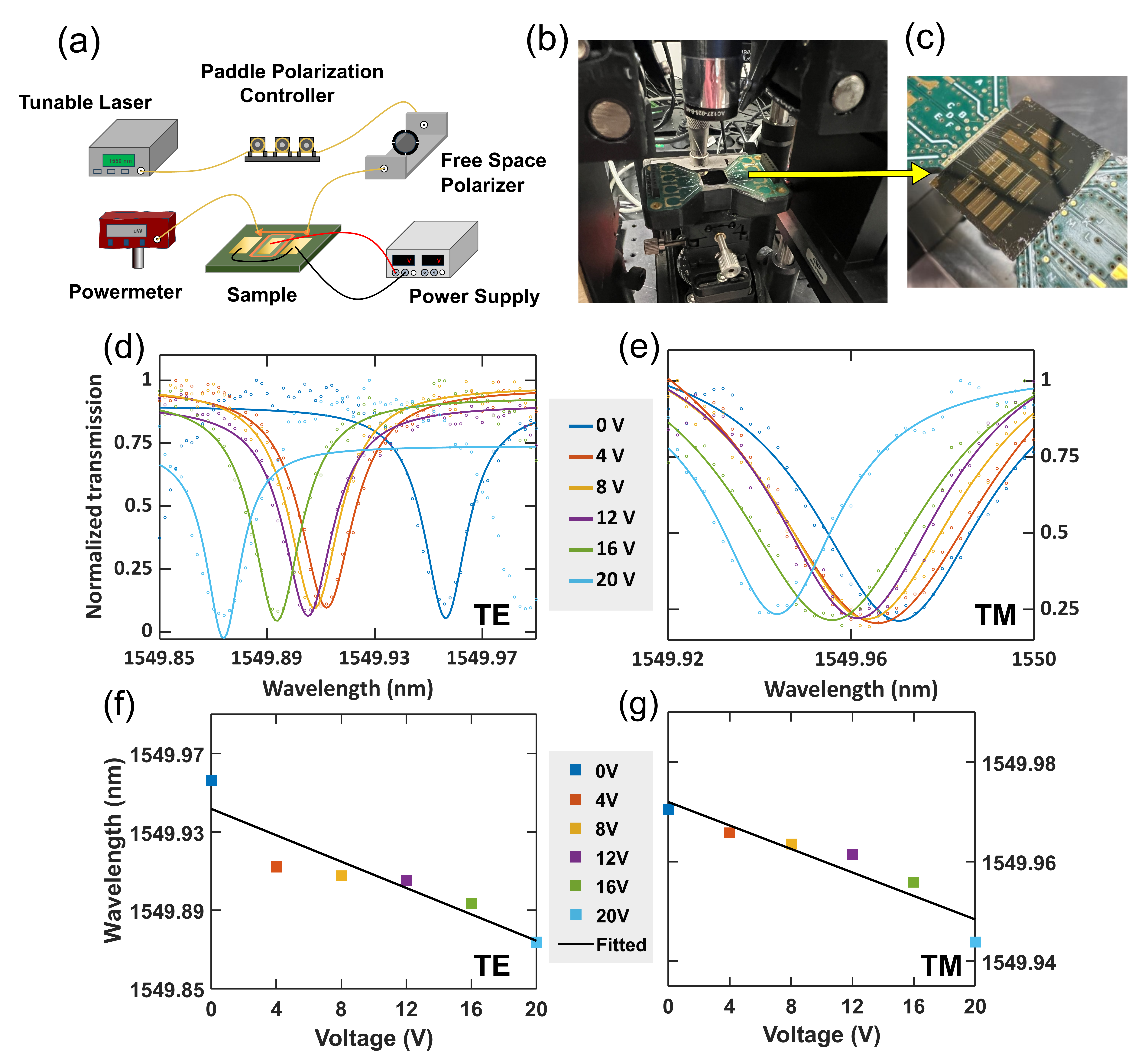}
    \caption{\label{fig:epsart} (a) Schematic diagram and (b) image of the characterization setup. (c) a-SiC/TFLN sample mounted and wire-bonded onto the PCB that connected to the power supply. Normalized transmission spectra of (d) TE mode and (e) TM mode propagated in the ring resonator when DC voltages applied (dots for raw data and solid lines for Lorentzian fittings). Resonance wavelength shifts with respect to DC voltage change (dots) and their linear fittings (solid lines) of (d) TE mode (3.4 pm/V) and (e) TM mode (1.2 pm/V). }
\end{figure*}

The electro-optics response of the a-SiC/LN platform is characterized and analyzed. Fig.4(a) depicts the characterization setup, starting from a tunable laser (Photonetics TUNICS-PRI 3642 HE 15), the light passes through a paddle polarization controller and a free-space polarizer (Thorlabs FBR-LPNIR). The free-space polarizer is used to select specific polarizations, while the paddle polarization controller aligns the input light to the selected polarization. The polarized light is then coupled by a polarization maintaining fiber into the a-SiC/LN device by the on-chip apodized grating coupler. The output light is coupled out from another apodized grating coupler and collected by an optical powermeter (Newport 818-NR). To measure the electro-optics response, the sample is mounted on a printed circuit board (PCB), shown in Fig.4(b). Afterwards, the on-chip Au electrodes are wire-bonded to the PCB (Fig.4(c)). A power supply is used to generate DC voltage signals, of which the ground and output ports are connected to the pads on PCB by probes. 

On the proposed heterogeneous a-SiC on TFLN platform, direct current (DC) driven Pockels effect is investigated to quantify the platform's tunability performance. Shown in the diagram in Fig.2(a), the racetrack ring resonator with Au electrodes is in push-pull configuration, with normal waveguides dimensions (260 nm $\times$ 800 nm) and gap $g$ = 4.8 $\mu m$. Fundamental TE and TM modes are excited and characterized when voltage sweeping is performed on the Au electrodes, shown in Fig.4(d) and (e). The center wavelengths of the resonance dips are extracted and plotted in Fig.4(f) and (g), together with linear fitted curves that represents EO tuning trends. Nonlinear change of EO coefficient can be observed, due to the DC bias drift of LiNbO$_3$ crystal happening during the long engaging time of wavelength sweeping for every voltage level\cite{salvestrini2011analysis,nagata2004dc,sun2020bias}. Based on the fitted EO response curves, the resonance tunability of 3.4 pm/V and 1.2 pm/V is realized, for TE mode and TM mode respectively. We calculated the half-wave voltage-length product and obtained 9.79 V$\cdot$cm for the TE mode and 27.5 V$\cdot$cm for the TM mode.. 

The difference between the simulation results and experimental results is believed to be related to the misalignment between racetrack longer arms and the LN crystal axis, also the mode profile is solved within the cross-section of the waveguides and considered to be consistently propagated in the third axis.

\section{Discussion and conclusion}
In conclusion, the a-SiC/LN heterogeneous photonic integrated platform is proposed and realized by a CMOS compatible fabrication processes at a temperature lower than 165$^\circ$C. Optical ring resonators are characterized and an intrinsic quality factor of 1.06 $\times$ 10$^5$ is measured. Pockels effect is evaluated on this platform by applying an electric field perpendicular to the mode propagation, consequently, 3.4 pm/V resonance tunability is achieved in the ring resonator. The applications of this a-SiC/LN platform can be extended to high-speed optical communications, programmable photonics, optical computing, nonlinear optics, and quantum optics. The optical loss can be further improved by optimizing the recipes in fabrication steps including deposition, lithography, etching and annealing. The waveguide dimensions (thickness and width) can be further engineered to significantly reduce the half-wave voltage-length production (for 260 nm narrow waveguides V$_{\pi}L$ is estimated to be 4.7 V$\cdot$cm, reducing a-SiC thickness can possibly push this value lower than 3 V$\cdot$cm). Feed-back loops and active corrections can be implemented to counter the bias point drift induced by long time applied DC-voltage. 
% expelling the hydrogen contents out of a-SiC film so that reduce the absorption. However, the annealing conditions must be carefully engineered, to make sure there is no damage to the LNOI substrate or undesirable diffusion taking place. 
In principle, the same heterogeneous integration can be applied on bulk lithium niobate substrate as well, utilizing higher nonlinear coefficients and stronger Pockels effect, compared to thin film LiNbO$_3$. Furthermore, the methods developed here and particularly low-temperature ICPCVD a-SiC constitute a promising route for heterogeneously integration with other ferroelectric substrates, such asbarium titanate and lead zirconate titanate. Therefore, with further development and optimization, the demonstrated platform holds great potentials for future heat-free tunable integrated photonics and represents a fundamental building block for second and third-order non-linearities on quantum communication and optical quantum computing.

\begin{acknowledgements}
The authors acknowledge the valuable comments and suggestions from Silvania. F. Pereira. Z.L. acknowledges the China Scholarship Council (CSC, 202206460012). N.S. and I. E. Z. acknowledge the funding from the NWO OTP COMB-O project (project 18757). I. E. Z. acknowledges funding from the European Union’s Horizon Europe research and innovation programme under grant agreement No. 101098717 (RESPITE project) and No.101099291 (fastMOT project).
\end{acknowledgements}

\section*{Author declarations}
\subsection*{Conflict of Interest}
The authors have no conflicts to disclose. 
\subsection*{Author Contributions}
\textbf{Zizheng Li:} Conceptualization (equal); Data curation (equal); Formal analysis (equal); Investigation (equal); Methodology (equal); Project administration (equal); Software (equal); Supervision (equal); Validation (equal); Visualization (equal); Writing – original draft (equal); Writing – review \& editing (equal). \textbf{Naresh Sharma:} Conceptualization (equal); Data curation (equal); Formal analysis (equal); Investigation (equal); Supervision (equal); Methodology (equal); Project administration (equal); Software (equal); Validation (equal); Visualization (equal); Writing – original draft (equal); Writing – review \& editing (equal). \textbf{Bruno Lopez-Rodriguez:} Conceptualization (equal); Data curation (equal); Formal analysis (equal); Investigation (equal); Methodology (equal); Supervision (equal); Validation (equal); Writing – original draft (equal); Writing – review \& editing (equal). \textbf{Roald van der Kolk:} Data curation (equal); Formal analysis (equal); Investigation (equal); Methodology (equal); Validation (equal); Writing – original draft (equal); Writing – review \& editing (equal). \textbf{Thomas Scholte:} Investigation (equal); Data curation (equal); Validation (equal); Writing – review \& editing (equal). \textbf{Hugo Voncken:} Data curation (equal); Investigation (equal); Writing – original draft (equal); Writing – review \& editing (equal). \textbf{Jasper van der Boom:} Data curation (equal); Writing – review \& editing (equal). 
% \textbf{Silvania F. Pereira:} Conceptualization (equal); Supervision (equal); Writing – original draft (equal); Writing – review \& editing (equal). 
\textbf{Simon Gröblacher:} Conceptualization (equal); Supervision (equal); Writing – original draft (equal); Writing – review \& editing (equal). \textbf{Iman Esmaeil Zadeh:} Conceptualization (equal); Funding acquisition (equal); Conceptualization (equal); Formal analysis (equal); Investigation (equal); Methodology (equal); Project administration (equal); Supervision (equal); Validation (equal); Visualization (equal); Writing – original draft (equal); Writing – review \& editing (equal).

\section*{Data Availability}

The data that support the findings of this study are available from the corresponding author upon reasonable request.

% \section*{REFERENCES}

\bibliography{aipsamp}% Produces the bibliography via BibTeX.

\end{document}